\def\conftype{0} 
\def\publish{1}

\ifcase 
\conftype  
\documentclass[sigconf, 10pt]{acmart}
\usepackage[english]{babel}
\usepackage{adjustbox}
\renewcommand\footnotetextcopyrightpermission[1]{} 
\setcopyright{none}
\acmDOI{}
\acmISBN{}
\acmPrice{}
\or 
\documentclass[letterpaper,twocolumn,10pt]{article}
\usepackage{usenix-2020-09}
\or
\documentclass[10pt, conference, letterpaper]{IEEEtran}
\fi

\usepackage{pifont}
\usepackage{tikz}
\usepackage[linesnumbered,ruled,vlined]{algorithm2e}
\usepackage[noend]{algpseudocode}
\usepackage{array}
\newcolumntype{x}[1]{>{\centering\arraybackslash\hspace{0pt}}p{#1}}
\usepackage{amsmath}
\usepackage[inline]{enumitem}
\usepackage{epsfig}
\usepackage{fancyvrb}
\usepackage{graphicx}
\usepackage{booktabs, tabularx}
\usepackage{makecell}
\usepackage[flushleft]{threeparttable}
\usepackage{latexsym, color}
\usepackage{listings}
\usepackage{tikz}
\usepackage{url}
\usepackage[font=normalsize,labelfont=bf]{caption}
\usepackage{hyperref}
\usepackage{cleveref}
\usepackage{tabularx}
\usepackage{subcaption}
\usepackage{xcolor}
\usepackage{multirow}
\usepackage{multicol}
\usepackage{float}
\usepackage{array}
\usepackage{algpseudocode}
\usepackage{setspace}
\usepackage{microtype}
 
\usepackage{amsmath,amsthm,amssymb,amsfonts}
\usepackage{minted}
\usepackage{float}
\usepackage{placeins}

\crefformat{figure}{Figure~#2#1#3}

\definecolor{commentgreen}{RGB}{2,112,10}
\definecolor{eminence}{RGB}{108,48,130}
\definecolor{weborange}{RGB}{255,165,0}
\definecolor{frenchplum}{RGB}{129,20,83}

\long\def\comment#1{}

\usepackage{cuted}
\usepackage{float}
\usepackage{xspace}
\usepackage{relsize}

\newcommand{\system}{\textscale{1.1}{\textit{SciNet}}\xspace}
\newcommand{\ie}{{\em i.e.}}
\newcommand{\eg}{{\em e.g.}}

\newcommand{\para}[1]{\noindent {\bf #1}}

\newcommand{\todo}[1]{{\color{black}#1}}

\begin{document}

\ifcase 
\conftype
\title[]{
Innovation Discovery System for Networking Research
}
\or
\title{}
\or
\title{

}
\fi


\author{
\ifcase\publish
  \or
     Mengrui Zhang$^{ 1\star}$, 
     Bang Huang$^{1 \star }$, 
     Yunxin Xu$^{1}$, 
     Haiying Huang$^{1}$, 
     Luxi Zhao$^{1}$, 
     Mochun Long$^{1}$, 
     Qingyu Song$^{1\ddagger}$,
    Qiao Xiang$^{1\ddagger}$,
    Xue Liu$^{2}$,
    Jiwu Shu$^{3}$, 
\\
 $^{1}$School of Informatics, Xiamen University, China\\
 $^{2}$School of Computer Science, McGill University, Montreal, Canada\\
 $^{3}$Minjiang University, China
  \fi	
}

\ifcase 
\conftype
\ifcase
\publish
\renewcommand{\shortauthors}{Anonymous authors}
\or
\renewcommand{\shortauthors}{Zhang et al.}
\fi
\fi

\ifcase \conftype  
\begin{abstract}
As networking systems become increasingly complex, achieving disruptive innovation grows more challenging. At the same time, recent progress in Large Language Models (LLMs) has shown strong potential for scientific hypothesis formation and idea generation. Nevertheless, applying LLMs effectively to networking research remains difficult for two main reasons: standalone LLMs tend to generate ideas by recombining existing solutions, and current open-source networking resources do not provide the structured, idea-level knowledge necessary for data-driven scientific discovery.

To bridge this gap, we present \system, a research idea generation system specifically designed for networking. \system is built upon three key components: (1) constructing a networking-oriented scientific discovery dataset from top-tier networking conferences, (2) simulating the human idea discovery workflow through problem setting, inspiration retrieval, and idea generation, and (3) developing an idea evaluation method that jointly measures novelty and practicality. Experimental results show that \system consistently produces practical and novel networking research ideas across multiple LLM backbones, and outperforms standalone LLM-based generation in overall idea quality.

\end{abstract}
\maketitle
\or 
\maketitle
\or
\maketitle
\fi
\footnotetext[1]{Mengrui Zhang and Bang Huang are co-primary authors. Qingyu Song and Qiao Xiang are co-corresponding authors.}



\section{Introduction}

Networking systems research has become increasingly complex, making disruptive innovation progressively more difficult. 
Meanwhile, the rapid development of LLMs has enabled notable progress in scientific hypothesis and idea generation. We refer to this line of work as \emph{LLM-based Scientific Discovery}. For example, AIScientist~\cite{aiscientist} uses LLMs with prompt templates to generate research directions. \textsc{AutoDiscovery}~\cite{autodiscovery} elicits prior and posterior beliefs about hypotheses via sampling, and uses surprisal as a reward within an MCTS procedure to balance exploration and exploitation in search of surprising discoveries. Other studies incorporate external knowledge to support idea generation, which we refer to as \emph{Data-driven Scientific Discovery}. For instance, \textsc{SciMon}~\cite{scimon} trains a language model~\cite{t5} on an open-source paper dataset~\cite{s2orc} to generate ideas and iteratively refine their novelty. Researchagent~\cite{researchagent} conducts literature reviews and integrates cross-domain knowledge with LLMs to generate and iteratively refine research ideas. However, these approaches are still not well suited to networking systems research, for two main reasons.

\para{LLMs tend to recombine existing solutions. }
Standalone LLM-based idea generation mainly draws from large-scale knowledge stored in model parameters, but this process is implicit and difficult to justify. In practice, the generated outputs are often recombinations of memorized patterns. For networking problems that require precise architectural design, general-purpose LLMs frequently produce ideas that are random combinations of existing designs rather than genuinely new mechanisms. Our experiments with standalone LLM-based idea generation also show that the generated ideas are often highly similar to existing solutions. 

\para{Open-source datasets lack networking knowledge. } 
In  bioinformatics and machine learning areas, large structured open-source literature resources are available~\cite{s2orc,spoke}. In the networking domain, however, although official organizations provide abstracts and full papers in PDF format~\cite{acm_dl,ieee_xplore,usenix_org,arxiv_org}, these resources are not structured for fine-grained, idea-level analysis. Existing third-party platforms that curate paper metadata (e.g., Semantic Scholar\cite {semanticscholar}, Open Alex~\cite{openalex}) either cover only a subset of networking papers or provide only abstract-level information. As a result, they are insufficient for \emph{Data-driven Scientific Discovery}. 

In this paper, we present \system for  networking research-specific idea generation. To our best knowledge, it is the first idea-generation system tailored to networking research. We build a scientific discovery dataset with domain-specific knowledge from premier networking conferences. Moreover, we propose to generate ideas with explicit scientific reasoning process, using knowledge-graph-based surveying and exploration with LLM-driven generation and iterative refinement.
Finally, we introduce an evaluation approach that jointly assesses idea novelty and practicality.
The three key designs (D1-D3) are detailed as follows.

\para{D1: Building a scientific discovery dataset for networking. } We collect papers published at two premier networking systems conferences
. We then use LLMs to extract each paper’s core information, unify heterogeneous structures and semantics, and construct a dataset rich in networking system design details with a consistent representation format.

\para{D2: Scientific reasoning-based idea generation.} 
We mimic the scientific reasoning process for idea generation, spanning research surveying, inspirational exploration, and idea development. Specifically, we use knowledge-graph search to emulate surveying and exploration, then combine the retrieved knowledge with an LLM to generate candidate ideas, select the one with the lowest similarity to existing solutions, and finally prompt the LLM to iteratively refine it.

\para{D3: Idea evaluation approach. }
To comprehensively evaluate the novelty and practicality of ideas generated by \system, we restrict both the input dataset and LLM resources to data available up to 2024, and then use them for idea generation. We subsequently compare each generated idea against two method sets: methods published before 2025 and methods published after 2024. Lower similarity to pre-2025 methods indicates higher novelty, while higher similarity to post-2024 methods indicates better practicality.

\todo{\para{Evaluation Results. } We implement a prototype of \system and evaluate the quality of the generated ideas, and the results show consistent effectiveness across different LLMs and research domains. We also conduct an ablation study, which demonstrate the effectiveness of \system's design.}

\vspace{-5pt}
\section{Overview}
\begin{figure}[t]
    \vspace{-10pt}
    \centering
    \includegraphics[width=\linewidth]{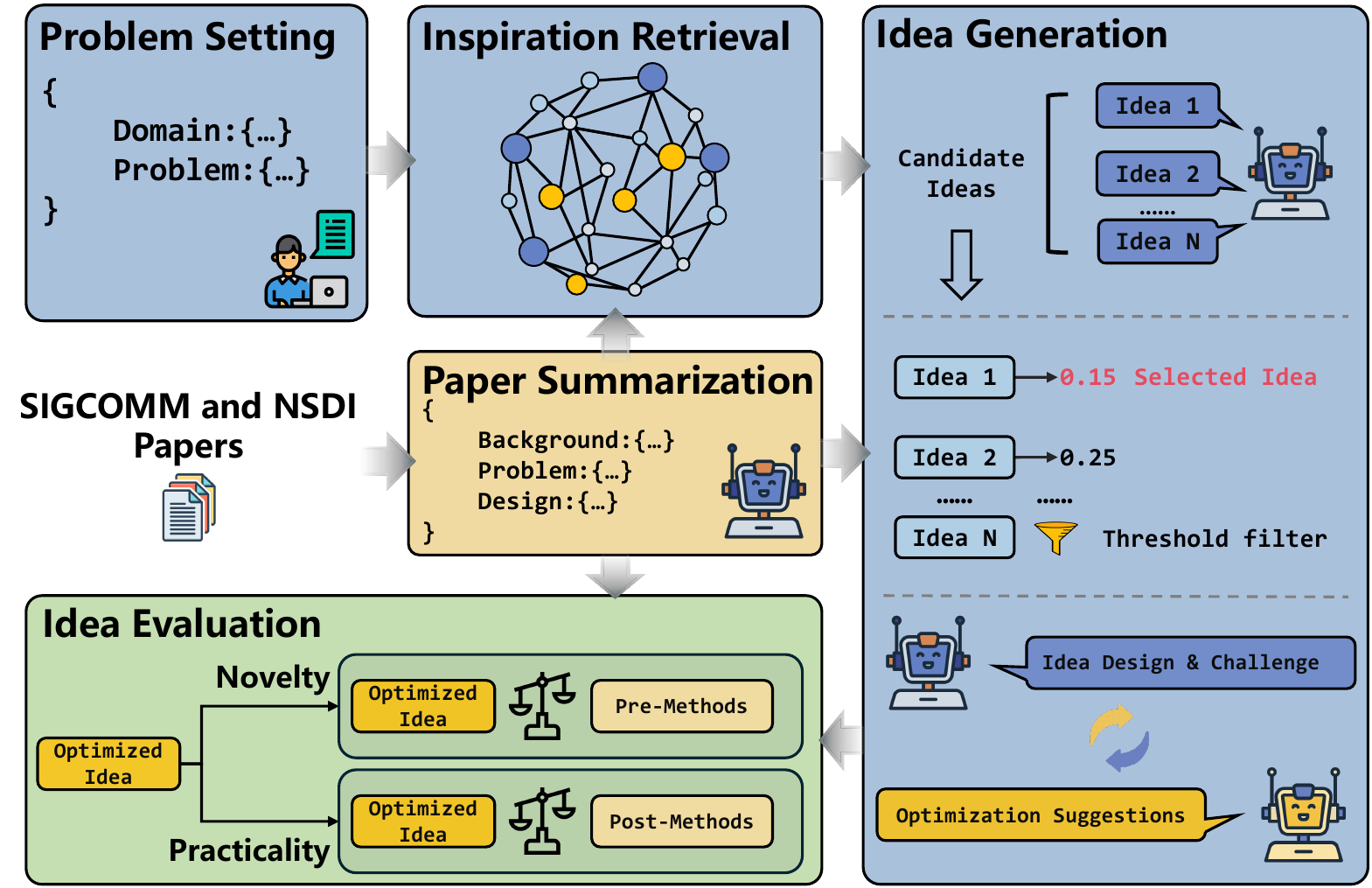}
    \caption{
    Framework: \system summarizes SIGCOMM\cite{sigcomm} and NSDI\cite{nsdi} papers into a structured dataset, builds knowledge graphs, retrieves methods for a user-specified domain and problem, generates and iteratively refines ideas with an LLM, and evaluates their novelty and practicality against the dataset.
    }
    \label{fig:framework}
    \vspace{-15pt}
\end{figure}

This section overviews \system. As shown in Figure~\ref{fig:framework}, its five components are organized into three parts matching our key designs: sand-yellow for Design D1, muted-blue for Design D2, and grass-green for Design D3.

\para{Paper Summarization. }
As shown in the middle-left part of Figure~\ref{fig:framework}, we collect papers published at ACM SIGCOMM~\cite{sigcomm} and USENIX NSDI~\cite{nsdi} from 2021 to 2025, and then use LLMs to summarize and unify each paper into three structured fields: $Background$, $Problem$, and $Design$. This step provides LLMs with detailed yet concise, semantically consistent, and structured networking domain knowledge for idea generation. Compared with raw PDFs (which are unstructured and semantically heterogeneous) or abstracts (which often omit implementation details), this representation better preserves system-level design information.

\para{Problem Setting. }
This step marks the beginning of idea discovery. Users first specify a research domain and a research problem, which substantially narrows the search space for both research surveying and inspiration exploration.

\para{Inspiration Retrieval. }
After the research domain and problem are specified, we mimic the processes of research surveying and inspiration exploration.

For research surveying, a naive strategy is to sequentially search the paper-summary dataset and retrieve related papers and methods. Although this approach may return relevant content, the results are often less explainable and are not fundamentally different from standard keyword search on platforms such as Google Scholar~\cite{googlescholar}. Instead of directly scanning papers one by one, we build a knowledge graph, called the \emph{paper graph}, to capture logical connections within and across papers. We then perform search over this graph to obtain explainable results about which existing papers and methods are related to the given research domain and problem.

For inspiration exploration, prior work shows that novelty tends to increase when ideas deviate further from established norms and standards ~\cite{surprising}. A naive strategy is therefore to randomly combine ideas from different domains, but this often leads LLMs to generate impractical or infeasible ideas. To address this issue, we use citation information to extend existing methods toward potentially relevant methods in other domains, and incorporate partial synthetic information to increase diversity. Based on these signals, we build another knowledge graph, called the \emph{citation graph}. We then search this graph to retrieve candidate inspirational methods.

\para{Idea Generation. }
After obtaining the research-survey and inspiration-exploration results, we combine them with the research domain and problem, and then use an LLM to generate candidate ideas (along with their technical challenges). We then compare each candidate idea with existing methods in the paper-summary dataset and select the least similar one as the initial idea. To further improve this idea with richer technical details, we prompt an LLM to provide optimization suggestions for addressing its technical challenges. We then iteratively refine the idea based on these suggestions, stopping when either the maximum number of iterations is reached or the LLM judges the idea to be mature.

\para{Idea Evaluation. }
Prior work\cite{scimon, ai-ideas, grapheval,zhang2026opennovelty} has mainly focused on evaluating novelty, while practicality is often validated through full-system implementation and experiments\cite{metamuse,alphaevolve}. In networking research, however, such implementation-heavy validation is costly and difficult to complete within a short cycle, even with LLM assistance\cite{xiang2023toward,repllm}. To evaluate both novelty and practicality in a scalable way, we adopt a time-split protocol. We partition papers into pre- and post-period sets, use only pre-period resources to build knowledge graphs and generate ideas, and align the LLM setting with the same temporal boundary. We then measure novelty against pre-period methods and practicality against post-period methods. This design enables a unified evaluation of both dimensions without requiring end-to-end prototype development for every generated idea.


The next three sections describe the construction of the paper-summary dataset and the knowledge graphs for research surveying and inspiration exploration ( \S~\ref{sec:dataset}), present the scientific reasoning-based idea generation process (\S~\ref{sec:framework}), and the idea evaluation approach (\S~\ref{sec:idea-evaluation}).

\begin{figure*}[t]
    \vspace{-10pt}
    \centering
    \includegraphics[width=\linewidth]{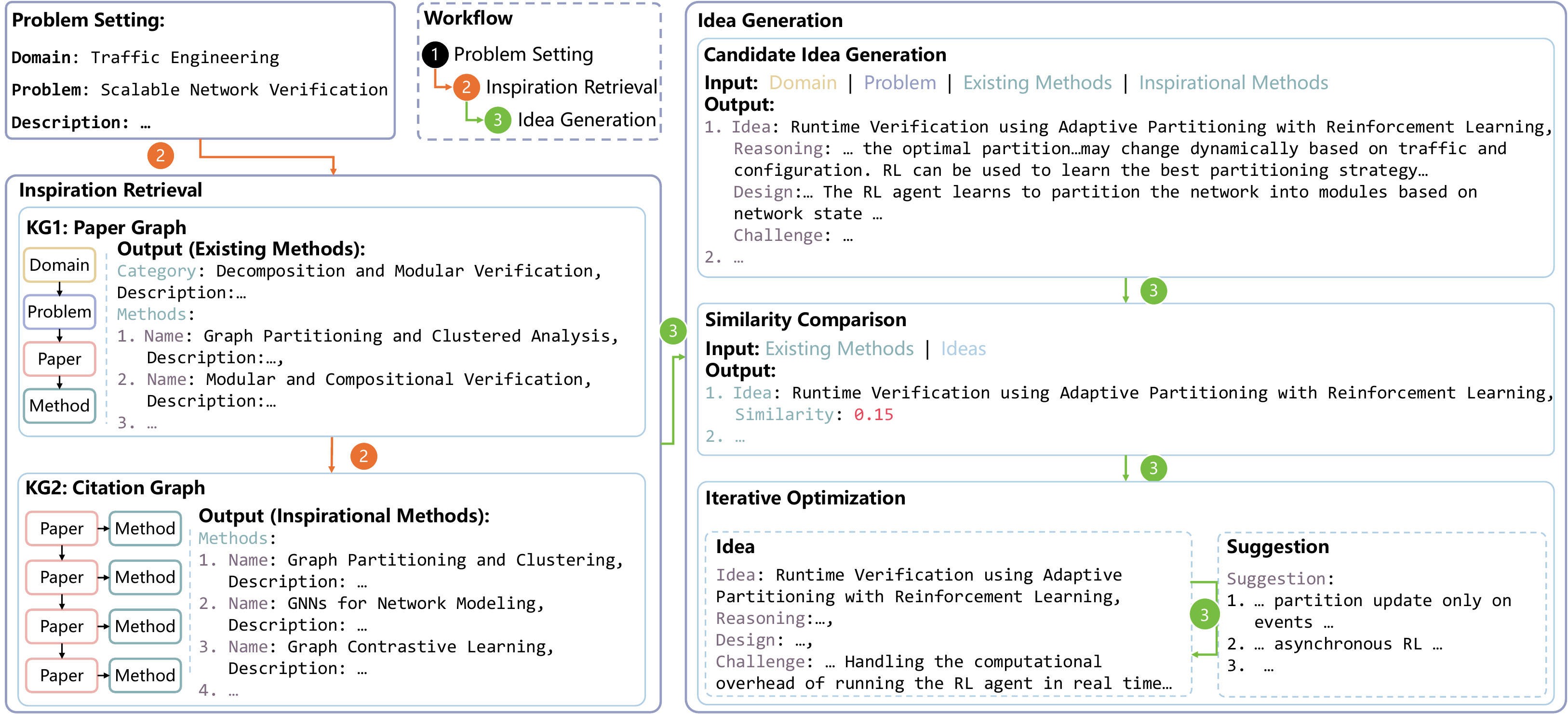}
    \vspace{-15pt}
    \caption{Workflow of Idea Discovery
    }
    \label{fig:workflow}
    \vspace{-10pt}
\end{figure*}

\section{Dataset $\&$ Knowledge Graph Construction}
\label{sec:dataset}
In this section, we describe how the summarized paper dataset is constructed (\S~\ref{sec:dataset-ps}) and how two knowledge graphs are built: the paper graph (\S~\ref{sec:dataset-pg}) and the citation graph (\S~\ref{sec:dataset-cg}).

\subsection{Paper Summary Dataset}
\label{sec:dataset-ps}

To build a high-quality dataset for idea generation, we collect papers published at two premier networking systems conferences, ACM SIGCOMM\cite{sigcomm} and USENIX NSDI\cite{nsdi}, from 2021 to 2025. The resulting dataset contains 743 papers. The summarization pipeline is divided into two steps.

\para{Semantic Cleaning. }
We first use LLMs to identify the research domain of each raw PDF paper. Because papers may describe the same domain at different levels of granularity or with different terminology, LLM outputs can vary across papers. For example, \emph{control-plane verification} and \emph{data-plane verification} can be merged into a unified domain, \emph{network verification}. After collecting the initial domain labels for all papers, we use \todo{LLMs} to merge semantically equivalent domains into standardized domain categories. In our dataset, $178$ raw domain labels are consolidated into $50$ unified domains.

\para{Paper Summarization. }
We then use \todo{LLMs} to summarize each paper into three structured components: $Background$ (research context and motivation), $Problem$ (core limitations of existing work and the technical challenges addressed), and $Design$ (the key method and system architecture proposed). We finally normalize these summaries into JSON format.




\subsection{Paper Graph}
\label{sec:dataset-pg}
To explicitly capture logical connections within and across papers, we construct a structured paper-level knowledge graph based on the paper-summary dataset. As shown in the left part of KG1 in Figure~\ref{fig:workflow}, the graph contains four entity types: $Domain$, $Problem$, $Paper$, and $Method$. For each paper summary, $Domain$ is extracted from $Background$, $Problem$ is extracted from $Problem$, $Paper$ is the paper title, and $Method$ is extracted from $Design$.

We then define three directed relationship types: \emph{<Domain, has, Problem>}, where a research domain contains specific problems; \emph{<Problem, is solved by, Paper>}, where a paper solves a specific problem; and \emph{<Paper, uses, Method>}, where a paper uses a specific key method or system architecture.

We then use GraphRAG \cite{graphrag} to construct the paper graph. After construction, the graph enables us to clearly identify which papers address which problems in which domains, as well as which methods are used in each paper.


\subsection{Citation Graph}
\label{sec:dataset-cg}
To extend existing solutions toward potentially relevant methods in other domains, we extract references from each paper (ACM SIGCOMM\cite{sigcomm} and USENIX NSDI\cite{nsdi}, 2021--2025) using MinerU \cite{mineru}. Since each paper contains approximately \todo{20--50} references, we first filter out non-article references and then randomly sample $k=20$ references per paper. We use Semantic Scholar \cite{semanticscholar} to retrieve the abstract of each sampled reference, and then use \todo{LLMs} to summarize its methods. We further randomly inject 5 synthetic citation links for each paper to increase transfer diversity while preserving practical feasibility.

Using the citation and method information, we construct the citation graph. As shown in the left part of KG2 in Figure~\ref{fig:workflow}, the graph contains two entity types, $Paper$ and $Method$, and two relation types:\emph{<Paper, cites, Paper>}, where a paper cites another paper as a reference; and \emph{<Paper, uses, Method>}, where a paper uses a specific key method or system architecture.


We then use GraphRAG \cite{graphrag} to construct the citation graph. After constructing this knowledge graph, we can clearly trace how methods are transferred and extended across papers.


\begin{figure*}[t]
    \centering
    \begin{subfigure}{0.48\linewidth}
        \centering
        \includegraphics[width=\linewidth]{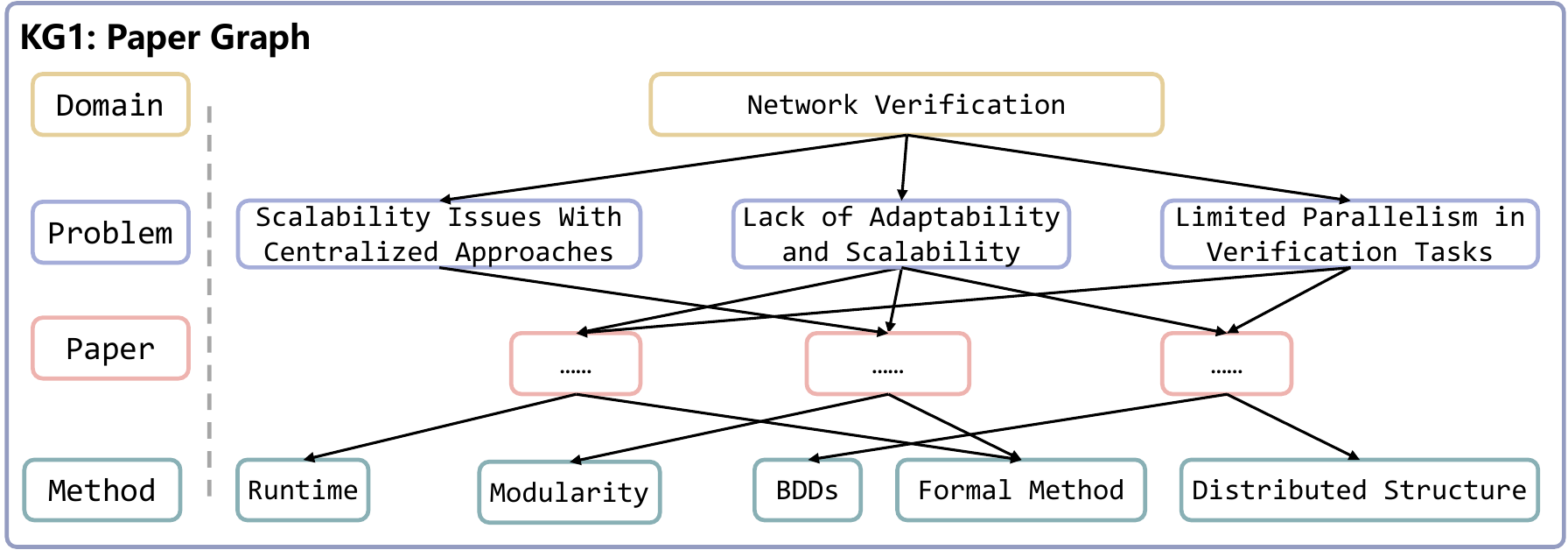}
        \caption{Subset of Paper Graph}
        \label{fig:kg1}
    \end{subfigure}
    \hfill 
    \begin{subfigure}{0.48\linewidth}
        \centering
        \includegraphics[width=\linewidth]{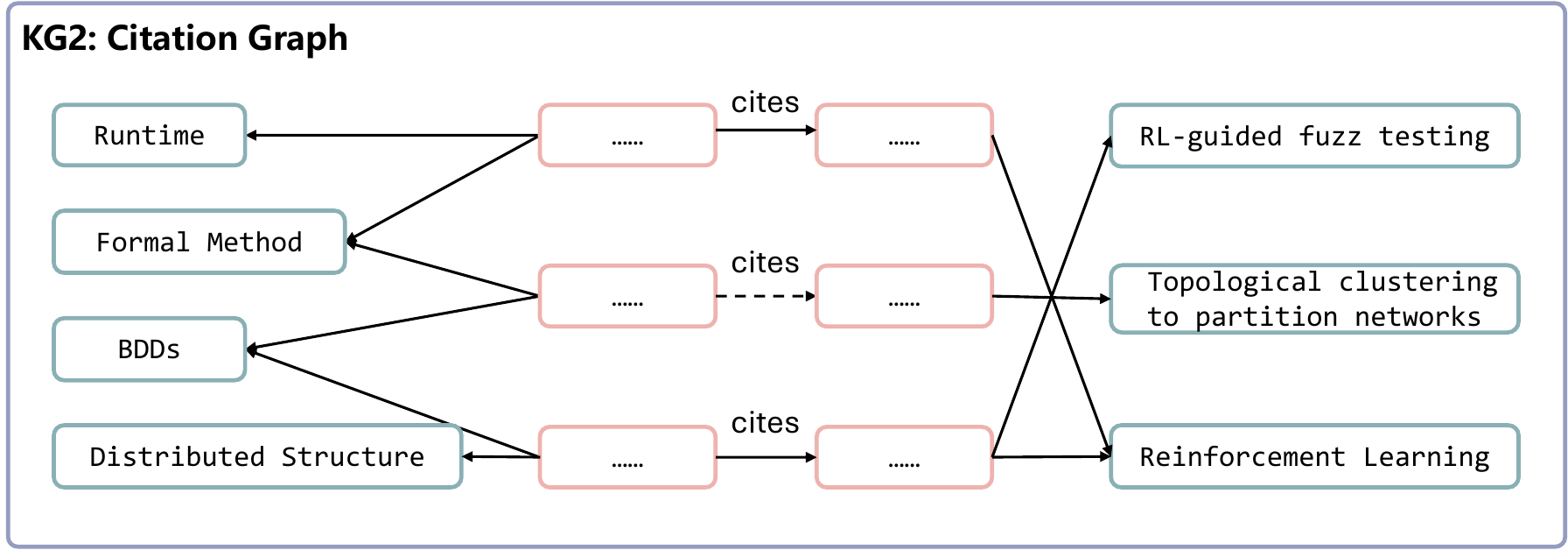}
        \caption{Subset of Citation Graph}
        \label{fig:kg2}
    \end{subfigure}
    \vspace{-5pt}
    \caption{
    Subsets of the \system Paper and Citation Graphs: (a) the Paper Graph links domains, problems, papers, and methods; (b) the Citation Graph links papers to cited papers and methods.
    }
    \label{fig:kg}
    \vspace{-10pt}
\end{figure*}

\section{Idea Discovery}
\label{sec:framework}
This section gives an example-driven walkthrough of \system's Idea Discovery workflow (Figures~\ref{fig:workflow} and~\ref{fig:kg}).

\subsection{Problem Setting}
\label{sec:framework-ps}
In the problem-setting stage, as shown in the upper-left part of Figure~\ref{fig:workflow}, \system takes the research background context $\mathcal{B}=\mathcal{D} \oplus \mathcal{P}$ as input, including: (1) a research domain $\mathcal{D}$ (\eg, \todo{network verification}) and (2) a research problem $\mathcal{P}$ with its description  (\eg, \todo{scalable network verification}). This input is then embedded into a knowledge-graph search prompt.
\subsection{Inspiration Retrieval}
\label{sec:framework-ir}
In the inspiration-retrieval stage, we retrieve information from two knowledge graphs: the Paper Graph and the Citation Graph. We adopt the global search function in GraphRAG\cite{graphrag}; unlike traditional graph search methods, global search generates answers by searching over all AI-generated community reports in a map-reduce manner. This strategy is particularly effective for questions that require a holistic understanding of the entire dataset \cite{graphrag}.

\para{Paper Graph. }
As shown in Figure~\ref{fig:kg1}, we present a subset of the Paper Graph for the example domain of \todo{network verification} and the problem of \todo{scalable network verification}. 
In Paper Graph, we retrieve existing methods $\mathcal{E}$ relevant to the research background defined in the problem-setting stage, such as \todo{"Graph Partitioning and Clustered Analysis''} and \todo{"Modular and Compositional Verification''} (Figure~\ref{fig:workflow}, KG1). We then categorize these methods into different orientations (\eg, Decomposition and Modular Verification).

\para{Citation Graph. }
After obtaining existing methods $\mathcal{E}_m$, we further retrieve inspirational methods $\mathcal{I}_m$ from the Citation Graph.
As shown in Figure~\ref{fig:kg2}, we present a subset of the Citation Graph for the example domain of \todo{network verification} and the problem of \todo{scalable network verification}.
For each existing method, we search for its potential connections and treat them as candidate inspirational methods, such as \todo{``Graph Partitioning and Clustering,'' ``GNNs for Network Modeling,'' and ``Graph Contrastive Learning''} (Figure~\ref{fig:workflow}, KG2).


\subsection{Idea Generation}
\label{sec:framework-ig}
In the idea-generation stage, we divide the process into three parts: Candidate Idea Generation, Similarity Comparison, and Iterative Optimization, as shown on the right side of Figure~\ref{fig:workflow}.

\para{Candidate Idea Generation. }
We first take the research background $\mathcal{B}$, existing methods $\mathcal{E}_m$, and inspirational methods $\mathcal{I}_m$ as input, \ie, $\mathcal{B}\oplus\mathcal{E}_m\oplus\mathcal{I}_m$. We then use \todo{LLMs} to generate $k=20$ candidate ideas $\mathcal{I}$ with technical challenges (\eg, \todo{``Runtime Verification using Adaptive Partitioning with Reinforcement Learning''} in Figure~\ref{fig:workflow}). Each idea includes a detailed design and step-by-step task decomposition. To ensure that each idea follows our reasoning principles 
, we enforce a structured output format that includes an explicit $Reasoning$ field, requiring the LLM to explain why the idea is likely to work.

\para{Similarity Comparison. }
\label{sec:framework-ig-sc}
We then take the existing methods $\mathcal{E}_m$ and candidate ideas $\mathcal{I}$ as input. We first embed both methods and ideas into vector representations using \todo{SPECTER\cite{specter2020cohan}}. For each idea $i \in \mathcal{I}$, we compute its cosine similarity with each existing method $e \in \mathcal{E}_m$. We then set a similarity threshold $t=0.8$. For each idea, we count the number of existing methods whose cosine similarity exceeds the threshold, denoted as $N_{>t}$, and denote the total number of existing methods as $N_{\mathcal{E}_m}$. We define the idea-selection score $S$ as
$
\frac{N_{>t}}{N_{\mathcal{E}_m}}
$
and select the idea with the minimum $S$ as the initial idea.

\para{Iterative Optimization. }
We then use the initial idea's technical challenges to optimize its technical details. We first prompt the LLM to provide suggestions for addressing these challenges (\eg, ``partition update only on events'' in Figure~\ref{fig:workflow}). We then iteratively refine the idea by incorporating these suggestions until it reaches the maximum iteration count (\todo{$i=10$}) or the LLM judges it to be a mature idea.

\section{Idea Evaluation Approach}
\label{sec:idea-evaluation}
In this section, we present an approach to evaluate whether \system can generate ideas that are both novel and practical.

To assess novelty, we compare generated ideas with existing methods based on similarity. To assess practicality, we leverage the observation that system papers with working prototypes generally reflect practically realizable designs; therefore, higher similarity to such methods can indicate higher practicality. However, high similarity may also imply limited novelty. To balance these two objectives, we adopt a time-split protocol. 
First, we split the paper-summary dataset into two subsets: papers from 2021--2024 (pre-dataset) and papers from 2025 (post-dataset). We rebuild the Paper Graph and Citation Graph using only the pre-dataset (pre-graphs), and use these pre-graphs together with LLM versions released before 2025 to generate ideas.

To evaluate novelty, we follow the process in \S~\ref{sec:framework-ig-sc} but compare only against existing methods in the pre-dataset ($\mathcal{E}_{pre}$). The novelty score $N_S$ is defined as:
\vspace{-5pt}
$$ N_S = 1 - \frac{N_{>t}}{N_{\mathcal{E}_{pre}}}$$
where $N_{>t}$ denotes the number of methods in $\mathcal{E}_{pre}$ whose cosine similarity with the generated idea exceeds the threshold $t$. A higher $N_S$ indicates a more novel idea.

To evaluate practicality, we consider the maximum cosine similarity between the generated idea and the methods in the post-dataset ($\mathcal{E}_{post}$). The practicality score $P_S$ is defined as:
\vspace{-10pt}
$$P_S = \max_{e \in \mathcal{E}_{post}} \; \text{cosine\_sim}(i, e)$$
where $i$ denotes the generated idea and $e$ denotes a method in $\mathcal{E}_{post}$. A higher $P_S$ indicates that the idea is closer to methods that appear later in the literature and is therefore more likely to be practical.

Finally, we combine novelty and practicality using the harmonic mean, and define the final idea quality score $I_S$ as:
$$I_S = \frac{2 \times N_S \times P_S}{N_S + P_S}
$$
This formulation encourages ideas that achieve both high novelty and high practicality, while penalizing cases where one metric is significantly lower than the other.

\begin{table*}[t]
\centering
\resizebox{\textwidth}{!}{
\begin{tabular}{l|ccc|ccc|ccc}
\hline
Metric 
& \multicolumn{3}{c|}{Congestion Control} 
& \multicolumn{3}{c|}{Traffic Engineering} 
& \multicolumn{3}{c}{Network Verification} \\
\cline{2-10}
& Gemini-2.0-flash & GPT-5 & Qwen-plus-1220
& Gemini-2.0-flash & GPT-5 & Qwen-plus-1220
& Gemini-2.0-flash & GPT-5 & Qwen-plus-1220 \\
\hline

Novelty
& 0.966 & 0.954 & 0.968
& 0.980 & 0.947 & 0.980
& 0.977 & 0.972 & 0.987 \\

Practicality
& 0.857 & 0.838 & 0.828
& 0.834 & 0.837 & 0.843
& 0.833 & 0.833 & 0.830 \\

Overall Quality
& 0.908 & 0.892 & 0.892
& 0.900 & 0.888 & 0.906
& 0.899 & 0.894 & 0.902 \\

\hline
\end{tabular}
}
\caption{Comparison across models and network domains.}
\label{tab:comparison}
\vspace{-15pt}
\end{table*}

\begin{figure*}[t]
    \centering
    \begin{subfigure}{0.33\linewidth}
        \centering
        \includegraphics[width=\linewidth]{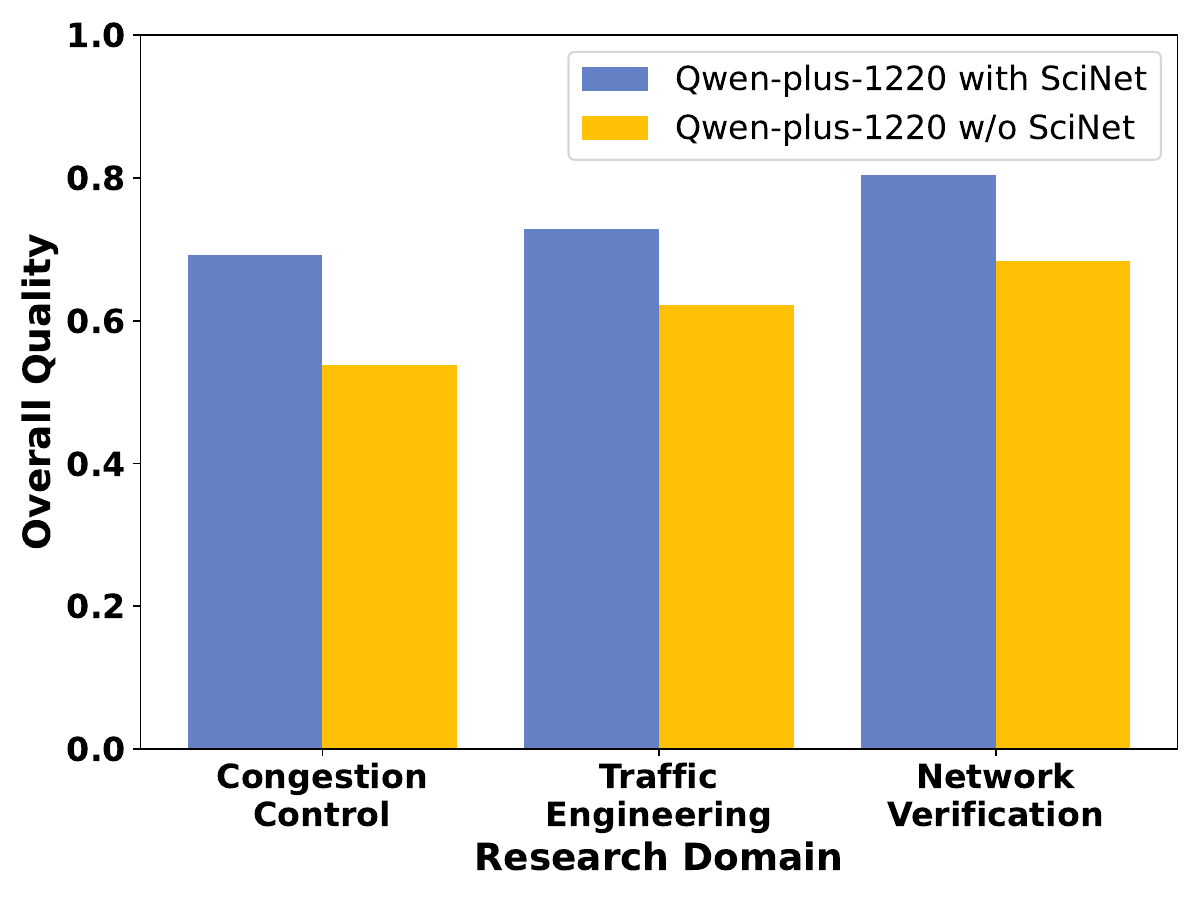}
        \caption{Performance on Qwen-plus-1220}
        \label{fig:qwen}
    \end{subfigure}
    \begin{subfigure}{0.33\linewidth}
        \centering
        \includegraphics[width=\linewidth]{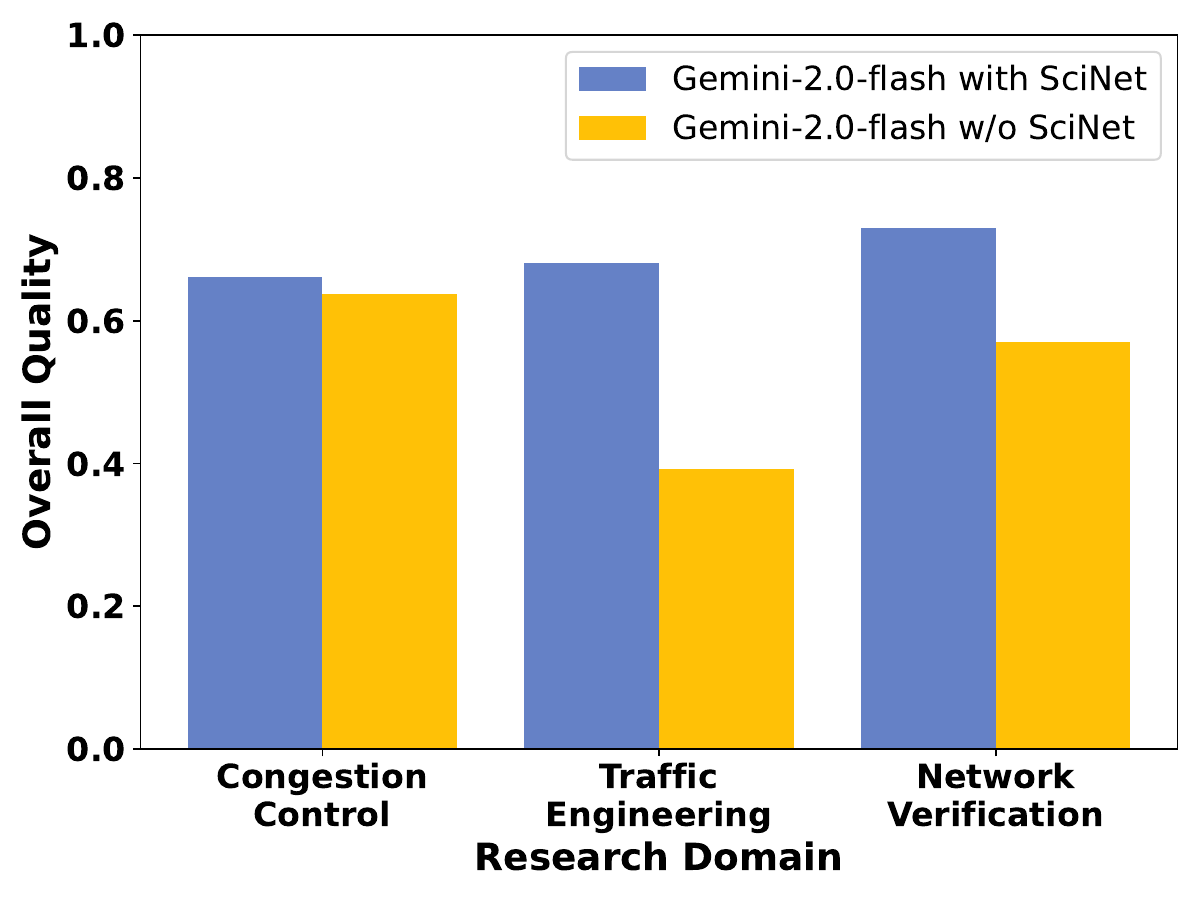}
        \caption{Performance on Gemini-2.0-flash}
        \label{fig:gemini}
    \end{subfigure}
    \begin{subfigure}{0.33\linewidth}
        \centering
        \includegraphics[width=\linewidth]{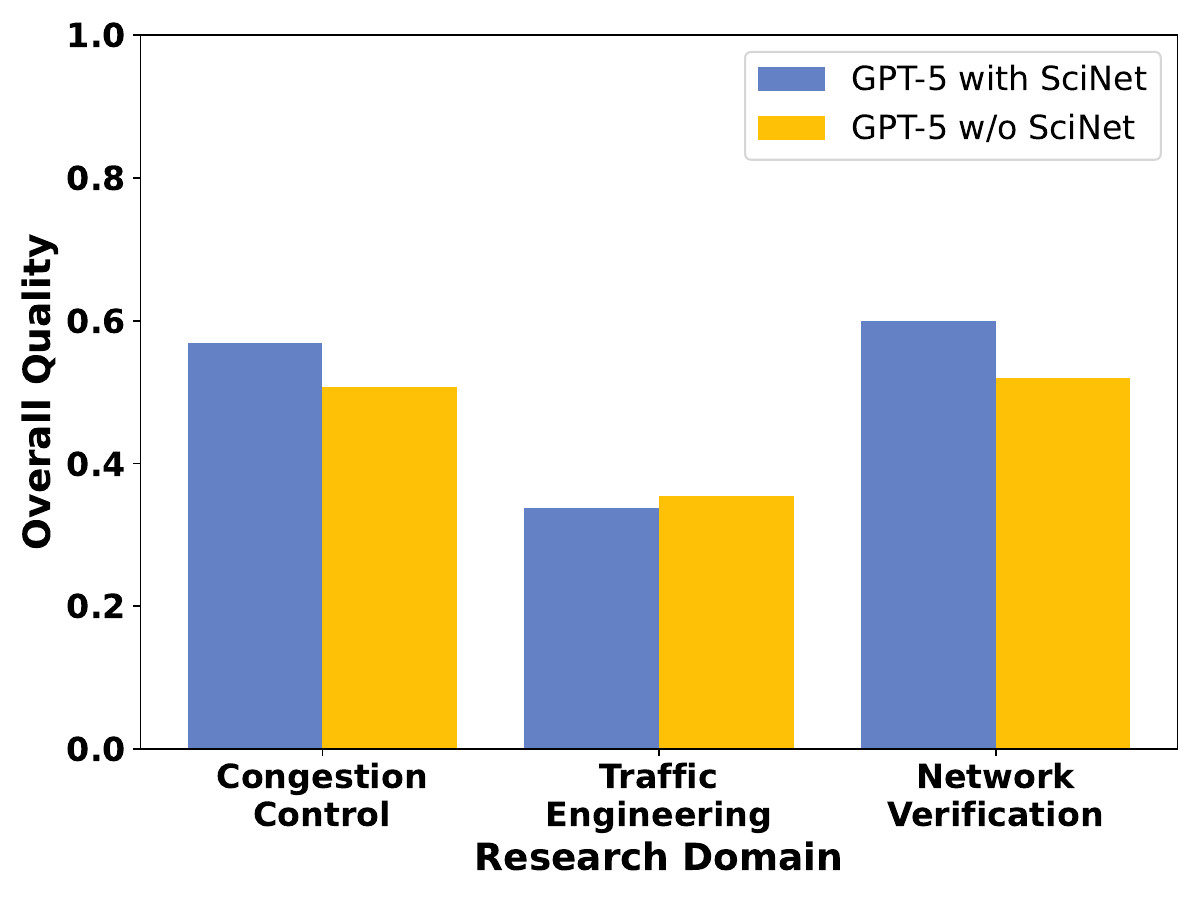}
        \caption{Performance on GPT-5}
        \label{fig:gpt}
    \end{subfigure}
    \vspace{-20pt}
    \caption{Ablation study of \system across different LLM backbones.}
    \label{fig:ablation}
    \vspace{-5pt}
\end{figure*}

\section{Evaluation}
\label{sec:eval}

We implement a prototype of \system and conduct extensive evaluations to investigate the following two research questions: (1) Can \system generate high-quality research ideas across different networking domains and language models? (2) How effective is the design of \system in improving the generation of ideas compared to standalone LLMs?

\subsection{Evaluation Setup}
\label{sec:evaluation-setup}

\noindent\hspace{1em} \textbf{1) Dataset:} To construct the evaluation dataset, we split the paper summary dataset introduced in \S~\ref{sec:dataset-ps} into two parts:

\para{Pre-Dataset (586 papers):} Papers published in 2021-2024. Ideas extracted from these papers are treated as existing ideas and serve as the source knowledge for idea generation.

\para{Post-Dataset (157 papers):} Papers published in 2025. These papers are reserved only for experiments; therefore, they serve as a natural benchmark for validating the quality of the generated ideas.

\textbf{2) Topics:} For evaluation, we select three representative networking areas to demonstrate the effectiveness of \system: congestion control, traffic engineering, and network verification.

\textbf{3) Metrics:} We evaluate the ideas along three dimensions: novelty, practicality, and overall quality. Specifically, we use an advanced scientific corpus embedding model \cite{specter2020cohan} to compute the similarity between generated ideas and those in our dataset. Then, we use the approach introduced in \S~\ref{sec:idea-evaluation} to calculate the scores.

\textbf{4) Models:} To ensure the validity of our evaluation, we select recent models with knowledge cutoffs up to 2024 from three representative large language model vendors. Specifically, we evaluate \textit{GPT-5} (knowledge cutoff: September 2024) from OpenAI \todo{\cite{chatgpt}}, \textit{Gemini-2.0-flash} (knowledge cutoff: August 2024) from Google \todo{\cite{gemini}}, and \textit{Qwen-plus-1220} (knowledge cutoff: November 2024) from Alibaba\todo{\cite{qwen}}.

\subsection{Idea Quality Comparison}
\label{sec:evaluation-comparison}

In this experiment, we aim to investigate the quality of ideas generated by \system across different domains when using different language models. The results are shown in Table~\ref{tab:comparison}. 

As observed from the table, the generated ideas exhibit strong novelty. Specifically, the novelty scores in the first row are all above 0.947, indicating that fewer than 6\% of the generated ideas are similar to existing ideas in the Pre-Dataset.
From the practicality line, it shows that the generated ideas consistently have similar counterparts in the Post-Dataset. This suggests that these ideas have strong potential to be realized and eventually developed into publishable research paper.
Moreover, when considering both novelty and practicality, the overall quality scores of the generated ideas are comparable across different domains and models. These results demonstrate that our system performs consistently well across multiple research areas and is not sensitive to the choice of underlying language model.

\subsection{Ablation Study}
\label{sec:evaluation-as}

To investigate the effectiveness of our design in \system, we conducted an ablation experiment, the results of which are shown in Figure~\ref{fig:ablation}.

Overall, across different models and research domains, applying our system consistently improves the quality of the generated ideas compared to the native outputs of the underlying language models. In particular, Figure~\ref{fig:gemini} shows that, when using Gemini, our system improves the quality score of ideas generated for the traffic engineering domain by 42\%. This improvement indicates that the Inspiration Retrieval stage of \system effectively identifies relevant existing methods and inspirational techniques, which are then incorporated through iterative refinement to guide the language model toward generating higher-quality ideas. In contrast, without the design, the ideas generated by standalone language models tend to overlap with a larger number of existing approaches, which leads to lower overall quality scores.

\section{Related Work} 

\para{Literature-Based Discovery. }
Literature-Based Discovery (LBD) aims to uncover new knowledge from existing literature in an automated manner \cite{swanson1986}. Its classical formulation traces back to Swanson’s ABC co-occurrence model \cite{abc}, which derives implicit relationships (\eg, ``A implies B'' and ``B implies C'') from explicit textual evidence. More recent work has employed link-prediction models \cite{sciagents,paperrobot,agatha} to discover scientific hypotheses as pairwise links between concepts. Recent studies have further leveraged LLMs \cite{scimon,aiscientist,researchagent,autodiscovery} to generate scientific hypotheses and novel ideas.

\para{Idea Assessment. }
Existing idea-assessment approaches can be roughly grouped into three categories: \emph{human assessment}\cite{scimon,aiscientist,si2024can,researchagent}, which evaluates novelty and practicality but is costly and hard to scale; \emph{experiment-based assessment}\cite{metamuse,alphaevolve}, which provides strong practicality evidence via prototypes or demos but requires long development cycles and high engineering cost; and \emph{automatic assessment}\cite{autodiscovery,zhang2026opennovelty,grapheval,sciagents}, which is efficient but either focuses mainly on novelty or risks hallucinations and factual uncertainty.


\section{Conclusion}
We present \system, an idea-generation system for networking research. \system integrates a scientific discovery dataset constructed from premier networking conferences, an explicit scientific reasoning process for  idea generation, and and a joint novelty-practicality evaluation approach. Experimental results show that \system consistently generates high-quality ideas across different LLM backbones.



\newpage

\ifcase \conftype
\bibliographystyle{ACM-Reference-Format}
\bibliography{ref}
\or
\bibliographystyle{plain}
\bibliography{ref}
\or
\bibliographystyle{IEEEtran}
\bibliography{ref}
\fi


\end{document}